\documentclass[a4paper,10pt]{iopart}

\usepackage{graphicx}
\usepackage[english]{babel}
\usepackage[utf8]{inputenc}
\usepackage{color}

\expandafter\let\csname equation*\endcsname=\relax
\expandafter\let\csname endequation*\endcsname=\relax
\usepackage{amsmath}
\usepackage{amsfonts}
\usepackage{amssymb}

\newcommand{\E}{\mathcal{E}}
\renewcommand{\S}{\mathcal{S}}
\newcommand{\vdE}{\langle\delta E^2\rangle}

\begin{document}

\title{A noise-immune cavity-assisted non-destructive detection for an optical lattice clock in the quantum regime}
\author{G. Vallet, E. Bookjans, U. Eismann\footnote{now with TOPTICA Photonics AG, Lochhamer Schlag 19, 82166 Graefelfing, Germany.}, S. Bilicki, R. Le Targat, and J. Lodewyck}
\address{SYRTE, Observatoire de Paris, PSL Research University, CNRS, Sorbonne Universit\'es, UPMC Univ. Paris 06, LNE, 61 avenue de l'Observatoire 75014 Paris, France}

\begin{abstract}
We present and implement a non-destructive detection scheme for the transition probability readout of an optical lattice clock. The scheme relies on a differential heterodyne  measurement of the dispersive properties of lattice-trapped atoms enhanced by a high finesse cavity. By design, this scheme offers a 1st order rejection of the technical noise sources, an enhanced signal-to-noise ratio, and an homogeneous atom-cavity coupling. We theoretically show that this scheme is optimal with respect to the photon shot noise limit. We experimentally realize this detection scheme in an operational strontium optical lattice clock. The resolution is on the order of a few atoms with a photon scattering rate low enough to keep the atoms trapped after detection. This scheme opens the door to various different interrogations protocols, which reduce the frequency instability, including atom recycling, zero-dead time clocks with a fast repetition rate, and sub quantum projection noise frequency stability.
\end{abstract}

\maketitle

\ioptwocol

\section{Introduction}

The performance of a clock is characterised by its level of systematic uncertainty, frequency stability, and reproducibility. As such, optical lattice clocks (OLCs) are at the forefront of frequency metrology. Not only have they reached systematic uncertainties in the low $10^{-18}$ in fractional frequency units~\cite{Nicholson2015,ushijima2015cryogenic}, but they have proven to be reliable and reproducible as it has been demonstrated by various measurement and comparison campaigns~\cite{Lodewyck2016,lisdat2016clock}. Albeit their success, the frequency stability of OLCs has not yet reached the quantum projection noise (QPN) limit, but is limited by the Dick effect, an aliasing effect between the frequency noise of the interrogation laser used to probe the clock transition (clock laser) and the cycle time of the experiment. OLCs have reached relative frequency stabilities in the low $10^{-16}/\sqrt{\tau}$~\cite{schioppo2016ultrastable, Nicholson2015}, with $\tau$ being the averaging time expressed in seconds. This value is well above the QPN, which is below $10^{-17}$ at one second for an experimentally reasonable number of atoms of over $10^{4}$. However, in order to achieve a QPN-limited frequency stability in an OLC, the technical noise sources, i.e. the frequency noise of the clock laser and/or the dead time between consecutive measurements, have to be reduced.

Currently, the most stable clock lasers are frequency-locked to ultra-stable cavities, which have reached stabilities on the order of $10^{-16}$, using e.g.  a long ultra-stable cavity \cite{Hafner:15, schioppo2016ultrastable, Nicholson2015} and a novel cryogenic single-crystal Si cavity \cite{Kessler2012}. To further push the frequency stability limit of the clock laser, new methods and techniques are being conceived, such as crystalline optical coatings \cite{Cole2013}, spectral hole burning in solid-state systems \cite{Chen2011,Leibrandt2013} and novel laser sources \cite{Meiser2009}. As an alternative to improving the clock laser frequency noise, the Dick effect can be reduced by minimising the dead time in the clock cycle. For this, a dead-time free clock can be designed~\cite{schioppo2016ultrastable, 1367-2630-12-6-065026}, but this requires two separate atomic ensembles. Another approach is to implement a detection system able to preserve the lattice-trapped atoms, and therefore substantially reduce the dead time in the clock cycle devoted to loading atoms~\cite{Lodewyck2009}. In this paper, we implement a cavity-assisted non-destructive detection technique in a Sr OLC. Its signal-to-noise ratio (SNR) also makes it suitable for quantum non-demolition measurements and therefore opens the path to correlated quantum states in OLCs and frequency stabilities beyond the QPN limit~\cite{hosten2016measurement, Schleier-Smith2010, PhysRevA.94.061601, Oblak2005, inoue2013unconditional, leroux2010orientation, Kuzmich1998,Wineland1994}.

\paragraph{Non-destructive detection methods in the dispersive regime.}

Dispersive measurement techniques are based on measuring the phase shift a probe beam acquires after passing through the atomic sample. They operate, in the far-detuned regime $\Delta \gg \Gamma$, with $\Delta$ being the detuning from resonance and $\Gamma$ the linewidth of the considered transition, and therefore have low photon scattering rates making them ideal for non-destructive measurements. In order to read out the induced phase shift, dispersive detection methods rely on interferometric techniques and require a stable phase reference. However, the advantage of a dispersive measurement is that all the information about the atomic population is gathered in the phase of the electromagnetic field of the mode of the probe beam, making it easy to collect and amplify with a cavity. In fluorescence based detection techniques, the atoms generally scatter in all directions, making it difficult to gather all the atomic information. In practice, this information loss is compensated by a large scattering rate, which leads to a large destructivity and consequently a substantial dead time in the clock cycle which needs to be devoted to trapping new atoms. The efficiency of dispersive measurement techniques has been exploited in various different cold and ultra-cold atom experiments and rely on various different interferometric techniques.  Examples include phase contrast imaging of BECs \cite{Andrews1996}, Mach-Zehnder interferometry~\cite{Oblak2005, Windpassinger2008}, heterodyne methods \cite{Bernon2011,Kohnen2011}, dual-color homodyning \cite{Lodewyck2009, Beguin2014}, cavity-based side-of-fringe methods~\cite{Schleier-Smith2010,Zhang2012}, and many more.

In this paper, we demonstrate a cavity-assisted non-destructive detection scheme based on a differential dispersive measurement. The scheme takes advantage of the parity of the dispersive atom-probe interaction such that it is to first order insensitive to technical noise sources, \emph{i.e.} the probing laser frequency noise and vibrations~\cite{Ye1998}. Similar to a previously implemented non-destructive detection scheme in a Sr clock using a Mach-Zehnder interferometer~\cite{Lodewyck2009}, the presented detection scheme measures the difference between the phase shifts induced on two modulation sidebands with opposite detuning with respect to the $^{1}$S$_{0}$--$^{1}$P$_{1}$ transition. Since the induced atomic phase shifts have opposing signs, the atomic signal adds, while the technical noise is in common mode and is subtracted out. A major improvement of this new scheme is the cavity design, which brings substantial advantages for detecting atoms in OLCs. The cavity is used for both the generation of the optical lattice potential (813 nm) and the non-destructive probe (461 nm). The local oscillator, used as a phase reference for the dispersive measurement, is a common-mode strong carrier uncoupled to the cavity. This geometry constrains the alignment between the optical lattice potential and the probe beam requiring them to be overlapped, and therefore provides the necessary stability, robustness, and reproducibility required for operational OLCs. More importantly, the fundamental SNR of the detection system is improved by the square root of the cavity finesse $\mathcal{F} = 1.6\times10^4$ through the increased interaction length between the atoms and the probe light. A common issue with cavity-assisted detection schemes is the fact that the probe beam forms a standing wave in which the atoms are only optimally coupled to the probe light at anti-nodes. Schemes where the trapping lattice and the probing lattice have commensurable pitches~\cite{Lee:14,hosten2016measurement} are not applicable to OLCs because the trapping lattice frequency is constrained by the magic wavelength for which the clock transition is unperturbed. By creating two lattices with opposite phases, the differential scheme introduced in this paper ensures an homogeneous coupling in the longitudinal direction, as also discussed in~\cite{PhysRevA.94.061601}.

\section{Theory}

The detection scheme considered in this paper relies on the dispersive properties of and atomic medium on a laser beam off-resonant from an atomic transition with wavelength $\lambda$. The phase shift $\phi^\text{at}$ induced by the atoms causes a displacement of the resonances of the cavity, which is detected in reflection by an heterodyne beatnote. This phase shift reads~\cite{Lodewyck2009}:
\begin{equation}
	\label{eq:phiat}
	\phi^\text{at} = \frac{N}{S}\frac{3\lambda^2}{2\pi}\frac{\Delta/\Gamma}{s + 4\Delta^2/\Gamma^2},
\end{equation}
where $N$ is the number of atoms in the fundamental clock state, $3\lambda^2/2\pi$ is the atomic cross-section, and $S$ is the cross-section of the cavity mode, averaged over the atomic ensemble. For an atomic sample of standard deviation $r_0$ probed by a Gaussian beam of waist $w_0$, $S = 2\pi (r_0^2 + w_0^2/4)$. $s$ is the saturation parameter, expressed by $s = P_c/(SI_\text{sat})$ where $P_c$ the power of the Gaussian beam, and $I_\text{sat} = \pi h c\Gamma/3\lambda^3$ is the saturation intensity. The measurement of $\phi^\text{at}$ therefore gives a measurement of $N$, from which the transition probability can be deduced using repumping methods~\cite{Lodewyck2009}.

\subsection{Noise insensitive detection}
\label{sec:theory}

\begin{figure}
\centering\includegraphics[width=\columnwidth]{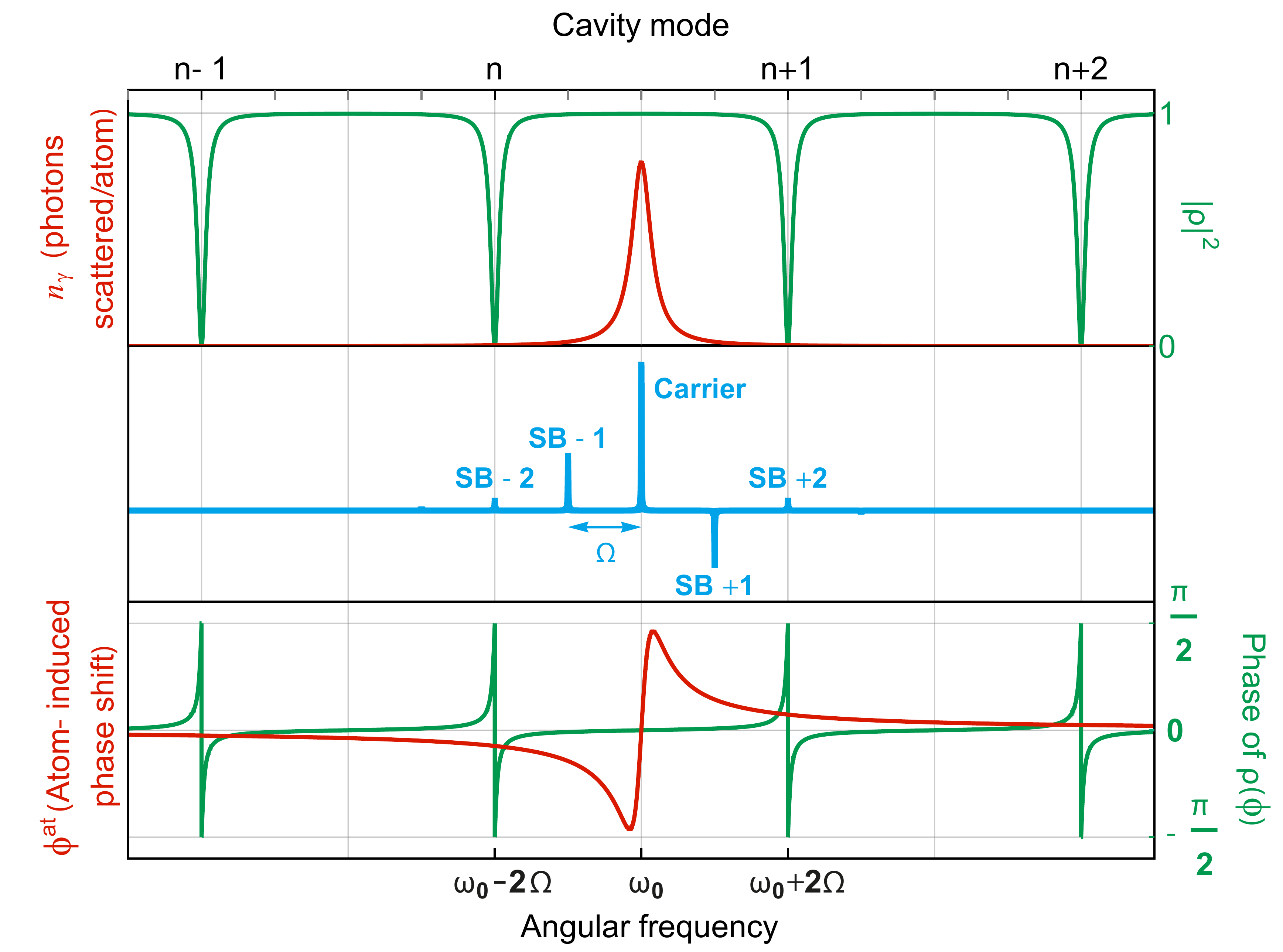}
\caption{\label{fig:scheme}Illustration of the frequencies used in the setup. \textbf{Top:} reflection coefficient of the cavity (green trace) and scattering rate of the atoms (red trace). \textbf{Middle:} in-phase quadrature of the phase modulated electromagnetic field incident on the cavity, showing the parity of modulation sidebands. Only the second order modulation sidebands are coupled in the cavity, while the carrier acts as an uncoupled strong local oscillator. \textbf{Bottom:} phase of the reflection coefficient of the cavity (green) and atom induced phase shift (red). The odd parity of the atomic phase shift induces a displacement of the frequency of the $n$ and $n+1$ resonances of the cavity with opposite sign, resulting in a net signal when demodulating at $2 \Omega = {\rm FSR}/2$. The displacement due to technical noise, however, induces a displacement with the same sign and therefore cancels out.}
\end{figure}

We consider a cavity of length $L = L_0 + \delta L$, where $L_0$ is the nominal cavity length and $\delta L$ are fluctuations around $L_0$. The cavity is illuminated with a laser beam of amplitude $|E_i|$, and angular frequency $\omega = \omega_0 + \delta\omega$, which is phase modulated with an angular frequency $\Omega = \Omega_0 + \delta \Omega$ and modulation depth $\beta$. The laser frequency and its modulation are tuned such that the second order modulation side-bands are resonant with two consecutive longitudinal modes of the cavity. The carrier is centred between these two modes:
\begin{equation}
	\frac{\Omega_0}{2\pi} = \frac 1 4 \frac{c}{2L_0}, \quad \text{and} \quad \frac{\omega_0}{2\pi} = \left(n + \frac 1 2\right) \frac{c}{2L_0},
\end{equation}
with $n$ being an integer. This configuration is schematically depicted in figure~\ref{fig:scheme}.

The one way phase shift $\phi_k$ of side-band $k$ inside the cavity reads:
\begin{align}
	\phi_k & = \frac{\omega + k\Omega}{c} L + \phi_k^\text{at} \\
	       & = \pi\left[n + \frac 1 2 + \frac k 4 + \left(n + \frac 1 2\right)\epsilon_\omega + \frac{k}{4}\epsilon_\Omega\right] + \phi_k^\text{at},\\
	       & \quad \text{with } \epsilon_x = \left(\frac{\delta x}{x_0} + \frac{\delta L}{L_0}\right),
\end{align}
where $\phi_k^\text{at} \equiv \phi^\text{at}(\omega + k\Omega)$ is the phase shift induced by the atoms on the sideband $k$. Because only the 2nd order sidebands are coupled to the cavity, the reflection coefficient of the cavity reads:
\begin{equation}
	\label{eq:rho}
	\rho_k \sim\left\{
	\begin{array}{ll}
	 -i\frac{2\mathcal{F}}{\pi}\phi_k     & \quad \text{for } \phi_k \simeq 0\ [\pi] \Leftrightarrow k = 2\ [4]\\
	 1 & \quad \text{otherwise}
	\end{array}
	\right.
\end{equation}
The electric field $E_d$ of the light collected by the detection photodiode can therefore be expressed as:
\begin{equation}
	\label{eq:sumE}
	E_d =  e^{i\omega t}\sum_{k=-\infty}^{+\infty} \E_k e^{ik\Omega t}, \quad \text{with } \E_k = \rho_k J_k(\beta) |E_i|,
\end{equation}
where $J_k$ are the Bessel functions of the first kind. This yields the AC photo-signal:
\begin{align}
	\S &= E_d E_d^* = \sum_{m > 0} \S_m, \text{ with}  \\
	\label{eq:Sm}
	\S_m &= 2 \Re \left[ e^{-im\Omega t} \sum_{k=-\infty}^{+\infty} \E_k \E_{k+m}^*  \right],
\end{align}
which is composed of harmonics with frequencies $m\Omega$, each derived from various beatnotes between sidebands separated by $m\Omega$.

Assuming that the modulation depth $\beta$ is small, the signal at frequency $\Omega$ reads:
\begin{equation}
	\label{eq:S1}
	\S_1 = 2 J_1(\beta) J_2(\beta)|E_i|^2  \frac{2\mathcal{F}}{\pi} (\pi (n + 1/2)\epsilon_\omega + \phi_{+2}^\text{at} + \phi_{-2}^\text{at})\sin\Omega t.
\end{equation}
It arises from the beating of $\pm 1$ order sidebands with $\pm 2$ order sidebands. If the frequency of the laser carrier is tuned close to the atomic resonance, $\phi_{+2}^\text{at} + \phi_{-2}^\text{at}$ is approximately zero, due to the odd parity of the atomic response. As a result, $\S_1$ is a measurement of $\epsilon_\omega$, \emph{i.e.} the relative laser frequency fluctuations with respect to the length of the cavity. This signal is thus used as an error signal to keep the laser frequency locked to the cavity resonance.

The signal at frequency $2\Omega$ reads:
\begin{align}
	\label{eq:S2}
	\S_2 & = \S_2^{\cos} \cos 2\Omega t + \S_2^{\sin} \sin 2\Omega t \quad \text{ with}\\
	\S_2^{\cos} & = - 2 J_1^2(\beta)|E_i|^2 \\
	\S_2^{\sin} & = 2 J_0(\beta) J_2(\beta) \frac{2\mathcal{F}}{\pi}   (\pi  \epsilon_\Omega + \phi_{+2}^\text{at} - \phi_{-2}^\text{at})|E_i|^2.
\end{align}
The $\S_2^{\sin}$ quadrature results from the beatnote between the carrier, used as a strong local oscillator, and the $\pm 2$ order sidebands coupled in the cavity. The even parity of these modulation sidebands make this signal insensitive to residual length fluctuations of the cavity and frequency fluctuations of the laser. Combined with the odd parity of the atomic response ($\phi_{+2}^\text{at} \simeq -\phi_{-2}^\text{at}$), the even parity of the $\pm 2$ order sidebands enables the detection of the atomic signal $\phi_{+2}^\text{at} - \phi_{-2}^\text{at}$. This signal is still sensitive to the relative mismatch between the cavity length and the modulation frequency $\epsilon_\Omega$, but can be made negligible through an independent lock of the cavity length. Furthermore, this latter sensitivity can be used to scale the photo-signal to a phase unit by purposely detuning $\Omega$.

\subsection{Shot noise analysis}

Shot noise is the fundamental quantum noise on the quadratures  of the electric field of coherent states. Its standard deviation only depends on the spatio-temporal geometry of the considered mode of the electric field. In the photodetection process, it is observed by its beatnote with an optical carrier.

\begin{figure}
	\centering \includegraphics[width=\columnwidth]{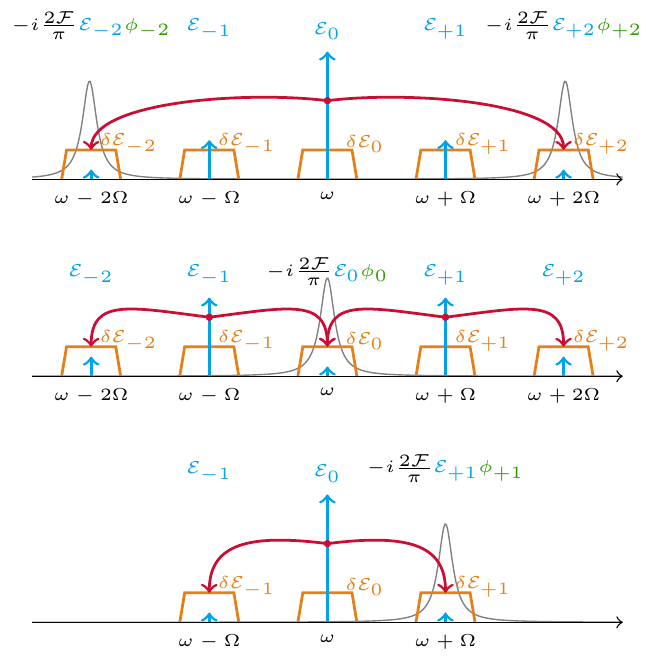}
	\caption{\label{fig:shotnoise}Illustration of the photon shot noise for different detection schemes. Blue arrows indicate the electric field reflected by the cavity, red arrows the betanotes which compose the demodulated signal, gray curves the cavity modes, and orange boxes the shot noise in the detection bandwidth. \textbf{Top:} detection scheme proposed in this paper. Its SNR is given by equation~(\ref{eq:SNR}). \textbf{Middle:} detection based on a PDH scheme. The first order sidebands are used as a local oscillator to detect the phase shift that the carrier coupled in the cavity experiences. Assuming that most of the detected optical power is in the first order sideband, equations~(\ref{eq:Sm}) and~(\ref{eq:dSm}) yield the SNR $\text{SNR}_\text{PDH} = \sqrt{2/3}\,(J_0(\beta)|E_i|/\sqrt{\vdE})(2\mathcal{F}/\pi)\phi_{0}^\text{at}$. It contains the same terms as equation~(\ref{eq:SNR}), except for a $\sqrt{2/3}$ factor which comes from the fact that while the PDH scheme samples the signal and shot noise at frequency $\omega$, it also samples the shot noise but no signal  at frequencies $\omega \pm \Omega$, hence a degraded SNR. \textbf{Bottom:} detection scheme in which a single optical sideband is coupled to the cavity. Its SNR reads $\text{SNR}_{+1} = \sqrt{1/2}\,(J_1(\beta)|E_i|/\sqrt{\vdE})(2\mathcal{F}/\pi)\phi_{+1}^\text{at}$. It also features the same structure as equation~(\ref{eq:SNR}), except for a $\sqrt{1/2}$ factor which comes from the shot noise at frequency $\omega - \Omega$.}
\end{figure}

Rewriting equation~(\ref{eq:sumE}) to include the shot noise gives:
\begin{equation}
	E_d = e^{i\omega t}\sum_{k=-\infty}^{+\infty} \left(\E_k + \delta\E_k \right) e^{ik\Omega t}.
\end{equation}
where $\delta \E_k$ is the shot noise on mode $\E_k$. Therefore, the shot noise of the mode $\S_m$ is:
\begin{equation}
	\label{eq:dSm}
	\delta \S_m = 2\sum_{k = - \infty}^{\infty}\Re \left[\left(\E_{k-m} \delta \E_{k}^* + \E_{k+m}^* \delta \E_{k} \right) e^{-im\Omega t}\right].
\end{equation}
Assuming a low phase modulation depth $\beta$, its component on the detection signal $\S_2^{\sin}$ hence reads:
\begin{equation}
	\delta \S_2^{\sin} =  2 J_0(\beta) |E_i|\Im\left[-\delta \E_2 + \delta \E_{-2}\right],
\end{equation}
which has a standard deviation:
\begin{equation}
	\sqrt{\langle\delta {\S_2^{\sin}}^2\rangle} =  2 \sqrt{2} J_0(\beta) |E_i| \sqrt{\vdE},
\end{equation}
given that the shot noise $\delta \E_2$ and $\delta \E_{-2}$ are uncorrelated and have an identical variance written $\vdE$. The shot noise limited SNR therefore reads:
\begin{equation}
	\label{eq:SNR}
	\text{SNR} = \frac{\S_2^{\sin}}{\sqrt{\langle\delta {\S_2^{\sin}}^2\rangle}} = \frac{\sqrt{2}J_2(\beta)|E_i|}{\sqrt{\vdE}}\frac{2\mathcal{F}}{\pi}\frac{\phi_{+2}^\text{at} - \phi_{-2}^\text{at}}{2}.
\end{equation}
In this equation, the $\sqrt{2}J_2(\beta)|E_i|$ factor is the amplitude of the electric field coupled into the cavity, and $2\mathcal{F}/\pi$ is the phase enhancement factor of the cavity, already seen in equation~(\ref{eq:rho}). The fact that the two second order sidebands contain an atomic signal is crucial for the optimality of the detection scheme, because both sidebands carry a signal in addition to the shot noise. To illustrate this fact, figure~\ref{fig:shotnoise} analyses other detection schemes which, in addition to being sensitive to cavity length and laser frequency fluctuations, feature a non-optimal SNR.

Finally, given that for a coherent state ${|E_i|}/{\sqrt{\vdE}} = 2\sqrt{N}$ with $N$ the number of photons in the mode, and taking into account the various optical losses in the detection process (the imperfect coupling into the cavity, the optical losses between the cavity and the photo-detector, and the quantum efficiency of the photo-detector, all combined into an effective efficiency $\eta$), equation~(\ref{eq:SNR}) can be reformulated as:
\begin{equation}
	\text{SNR} = \sqrt{\frac{4\eta P_c T \lambda}{hc}\frac{2\mathcal{F}}{\pi}}\frac{\phi_{+2}^\text{at} - \phi_{-2}^\text{at}}{2},
\end{equation}
where $T$ is the integration time, and $P_c $ is the total intra-cavity power held by both second order sidebands injected in the cavity. This expression is a factor $\sqrt{2\mathcal{F}/\pi}$ higher than the SNR of a dispersive detection scheme without a cavity~\cite{Lodewyck2009}.

\subsection{Non-destructivity}

We recall that, in the detection scheme described here, the frequency of the optical carrier of the detection laser is close to the frequency of the atomic resonance. Therefore, the detunings of the second order modulation sibebands are opposite in sign, and are written as $\pm\Delta$. It is important to remark that because these two sidebands are injected into two consecutive modes of the cavity, the interference patterns they each form have opposing phases close to the centre of the cavity where the atoms are located. Therefore, the total intra-cavity power experienced by the atoms is homogeneous along the longitudinal direction. As the phase shifts of the two modulation sidebands are jointly measured by the heterodyne scheme presented here, the coupling between the atoms and the cavity modes, as well as the photon scattering rate therefore do not depend on the position of the atoms along the axis of the cavity. This is a compelling feature of this detection scheme for OLCs, for which the trapping light has to be tuned to the magic wavelength, \emph{i.e.} to a frequency which is incommensurable with the frequency of the detection light.

The heating rate of the atoms in the cavity is characterised by the photon scattering rate per atom:
\begin{equation}
	n_\gamma = \frac{\Gamma}{2} \frac{s}{s + 4\Delta^2/\Gamma^2}.
\end{equation}
The SNR can consequently be expressed as:
\begin{equation}
	\text{SNR} = \frac{\phi_{+2}^\text{at} - \phi_{-2}^\text{at}}{\delta\phi},
\end{equation}
with
\begin{equation}
	\frac{1}{\delta\phi} = \sqrt{\frac{2\pi}{3} \frac{S}{\lambda^2}\eta n_\gamma T\frac{2\mathcal{F}}{\pi}}\sqrt{s + 4\Delta^2/\Gamma^2}
\end{equation}

Using the expression (\ref{eq:phiat}) for the atomic phase shift, the shot-noise limited SNR can be written in terms of the atom number resolution:
\begin{equation}
	\text{SNR} = \frac{N}{\delta N},
\end{equation}
with
\begin{equation}
	\label{eq:deltaN}
	\frac{1}{\delta N} = \sqrt{\frac{3}{2\pi} \frac{\lambda^2}{S}\eta n_\gamma T\frac{2\mathcal{F}}{\pi}}\sqrt{\frac{4\Delta^2/\Gamma^2}{s + 4\Delta^2/\Gamma^2}}.
\end{equation}
For $s \ll 4\Delta^2/\Gamma^2$, the SNR depends only an a few experimental parameters, namely the atomic cross-section $3\lambda^2/2\pi$, the cavity cross-section $S$, the finesse $\mathcal{F}$, the detection efficiency $\eta$, and the number of scattered photons per atom $n_\gamma T$. If the latter is sufficiently low to prevent the atoms from escaping the trapping potential, the detection exhibits ``classical'' non-destructivity. If the number of scattered photons is  significantly lower than one, the detection lies in the quantum regime~: it is able to weakly measure the atomic coherence. If the detection in the quantum regime can reach a detection noise smaller than the quantum projection noise, \emph{i.e.} $\delta N \ll \sqrt{N}$, spin-squeezed states of the atomic ensemble can be generated to overcome the quantum projection noise. A cavity with a large finesse therefore helps reach this quantum regime.

The incoming probe beam on the cavity is composed of two cavity-coupled, weak $\pm2$ modulation sidebands, but also of a strong carrier which acts as a local oscillator. Because a low destructivity implies a low power in the coupled sidebands, hence a low modulation depth, the destructivity of the non-cavity coupled but strong and carrier also has to be considered because it is on-resonance with the atoms. The destructivity of the carrier due to residual off-resonant coupling into the cavity can be mitigated by purposefully introducing a non-zero detuning between the carrier and the atomic resonance, but sufficiently small as compared to the modulation frequency in order to preserve the homogeneous coupling of the atoms in the cavity. This condition is in practice always met, because once the cavity is injected with the trapping light at the magic wavelength, the frequencies of the resonances at the detection wavelength form a discrete set fixed by the incommensurable ratio between these frequencies. Also, the detection time can be decreased while keeping the product $P_cT$ constant, allowing for a larger modulation depth.

\section{Experimental demonstration}

\subsection{Setup}

We experimentally implemented the detection scheme proposed in this paper on one of the two operational strontium optical lattice clocks available at SYRTE, named Sr B. This clock features a systematic uncertainty of $4\times10^{-17}$, which has been confirmed by a relative frequency ratio measurement between Sr B  and the other Sr clock (Sr 2), yielding
\begin{equation}
	\frac{\nu_{\text{Sr}_2} - \nu_{\text{Sr}_B}}{\nu_\text{Sr}} = (2.3 \pm 7.1)\times 10^{-17}
\end{equation}
The frequency stability of Sr B is $10^{-15}$ for an integration time of 1~s, measured though a comparison with Sr 2~\cite{Lodewyck2016}. The aim of the non-destructive detection described in this paper is to improve this frequency stability.

\begin{figure}
\center
\includegraphics[width=\linewidth]{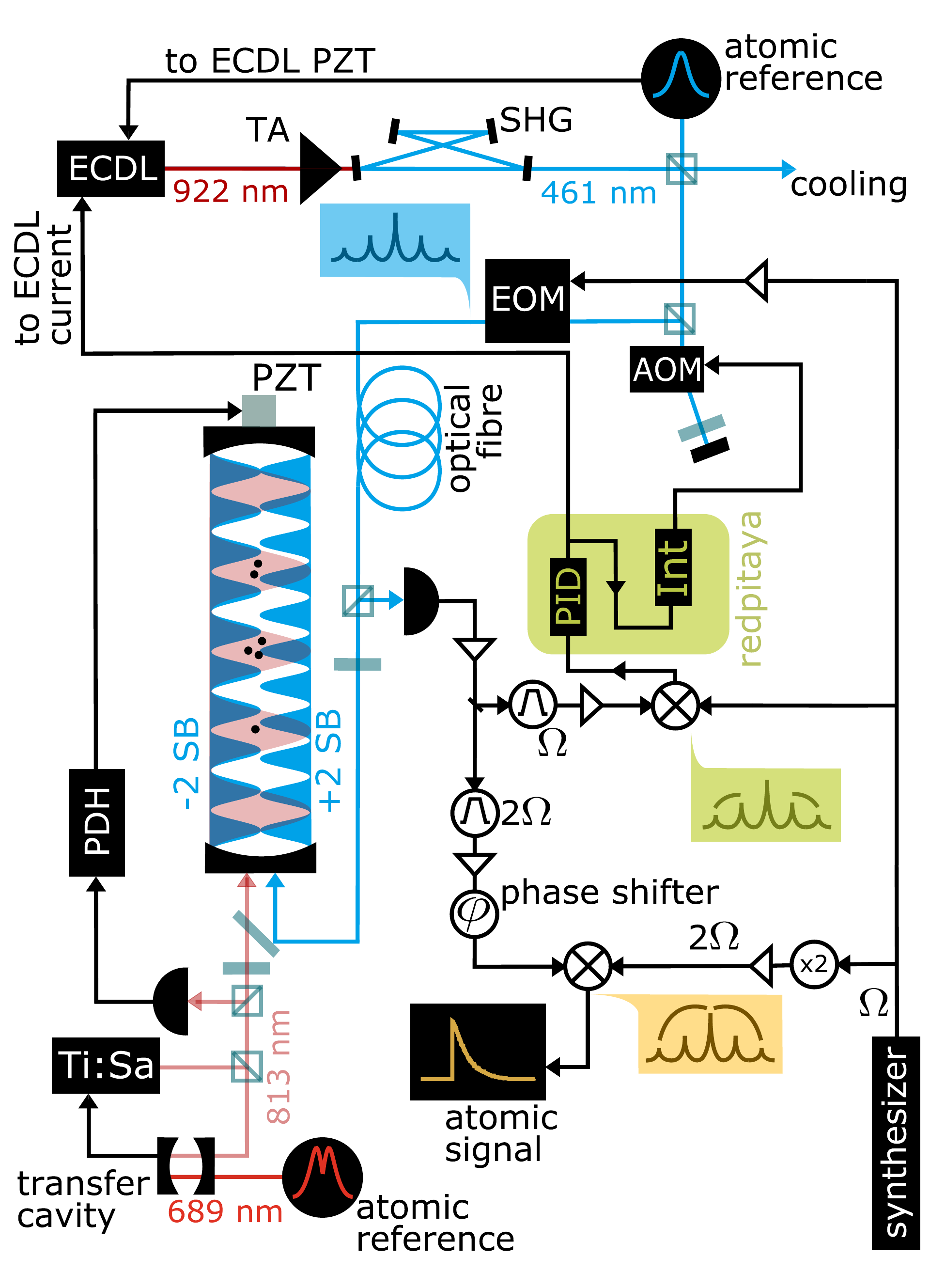}
\caption{\label{fig:expscheme}Schematic of the experimental setup. It shows the bi-chromatic high finesse cavity in which the atoms are trapped (magic wavelength at 813~nm) and probed (phase modulated laser at 461~nm). The second-order modulation sidebands form two out-of-phase standing waves in which the atoms are homogeneously coupled.}
\end{figure}

The description of the clock system can be found in~\cite{Lodewyck2016}. Here, we describe exclusively the parts of the experiment related to the detection scheme, shown on figure~\ref{fig:expscheme}.

The main part of the detection scheme is an in-vacuum bi-chromatic build-up cavity used for both the optical lattice confining the Sr atoms in the Lamb-Dicke regime and the enhancement of the atomic phase shift imprinted on the non-destructive probe light. The cavity has a finesse of 180 at the trapping magic wavelength $\lambda_{m}\simeq 813$~nm, which allows us to reach trapping depths as large as 2500~$E_{R,813}$, with $E_{R,813} = h^2/2m\lambda_m^2$ being the recoil energy of Sr atoms of mass $m$. The finesse at the non-destructive detection wavelength 461~nm, corresponding to the $^{1}$S$_{0}$--$^{1}$P$_{1}$ transition ($\Gamma = 2\pi\times 32$~MHz and $I_\text{sat}$ = 42.7~mW/cm$^2$), is $\mathcal{F} = 1.6\times 10^4$, which allows for a large enhancement of the detection signal. The two mirrors of the cavity are used as windows of the high vacuum system, and are adjusted by bellows to which piezo transducers (PZT) are attached. The length of the cavity is $L=41$~mm, corresponding to a free spectral range of 3.7~GHz. The radius of curvature of the mirrors is 25~mm, which gives a waist of 50~$\mu$m at 813~nm, and $w_0 = 38~\mu$m at 461~nm. Given that the typical transverse standard deviation of the atomic wavefunction is $r_0 = 10~\mu$m, the cavity cross-section is $S = 2.9\times10^3~\mu$m$^2$.

The 813~nm lattice light is provided by a titanium sapphire (Ti:Sa) laser. It is tuned to the magic wavelength $\lambda_{m}$ by a frequency-lock to an extended cavity diode laser (ECDL, not represented in figure~\ref{fig:expscheme}) referenced to a transfer cavity, itself locked to a 689~nm source directly referenced to the Sr $^1S_0$--$^3P_1$ transition by saturation spectroscopy. The length of the cavity is then locked to the magic wavelength lattice light via a standard Pound Drever Hall scheme acting on the PZT.

The 461~nm blue laser stems from a frequency doubled 922~nm ECDL. It is then dispatched to the first Sr cooling stage light system. Its frequency is locked to an atomic reference through a feedback loop onto the 922~nm ECDL PZT using a digital modulation lock. Part of the light is sent to the cavity for the non-destructive detection. Because the frequencies of both the 813~nm and 461~nm lasers are constrained, a double passed Acousto-Optic Modulator (AOM) is used to tune the frequency of the 461~nm light to the resonances of the cavity. The laser beam is then phase modulated by an Electro-Optic Modulator (EOM, provided by QuBig) at a frequency of one forth of the free spectral range, \emph{i.e.} $\Omega = 2\pi\times920$~MHz, in order to implement the heterodyne protocol described in section~\ref{sec:theory}. After being cleaned through an optical fibre, the spatial mode of the laser is coupled to the cavity with an efficiency of $\eta_\text{cpl} = 0.82$.

Despite the fact that the cavity length and the laser frequencies are all locked to absolute references, the frequency difference between the 461~nm light and the resonances of the cavity fluctuates by much more than the cavity line-width, on time scales of tens of microseconds. This is due to the high finesse of the cavity, which is two orders of magnitude larger at 461~nm than at 813~nm. To keep the probe laser on resonance with the cavity, it is therefore necessary to implement a fast locking scheme of the laser frequency to the cavity length. For this, we use the protocol described in section~\ref{sec:theory}. The reflected signal is collected by a fast photodiode (OSIO FCI-125G-006HRL, with a specified bandwidth of 1.25~GHz and a quantum efficiency $\eta_\text{det} = 0.46$) after an additional transmission efficiency $\eta_\text{loss} = 0.65$, yielding an overall efficiency $\eta = \eta_\text{cpl}\eta_\text{det}\eta_\text{loss} = 0.25$. The RF output signal of the photodiode is then split into two.

The first part is bandpass-filtered and demodulated at frequency $2\Omega$ in order to generate the atomic signal. This signal results from the beatnote of the reflected carrier acting as a strong local oscillator and the coupled second-order sidebands, and is given by equation~(\ref{eq:S2}). It is calibrated by applying a modulation on the modulation frequency $\Omega$, and the demodulation quadrature is adjusted in order to maximise the modulation signal using a mechanical phase shifter.

The second part is bandpass-filtered and demodulated at frequency $\Omega$. It contains the error signal given by equation~(\ref{eq:S1}) used to lock the 461~nm light frequency on the cavity. It results from the beatnote of the rejected first-order sidebands (local oscillator) and the coupled second-order sidebands. The servo loop uses a digital PID implemented on a FPGA based RedPitaya platform acting on the current of the 922~nm ECDL with a bandwidth larger than 1~MHz. This digital implementation also enables us to conditionally engage the integration stage of the PID with a TTL signal to accommodate with the fact that the blue light is turned off during the clock interrogation time. The correction signal is kept at zero by a second stage integrator acting on the AOM frequency. The aim of this lock is to keep the cavity in the linear regime for which the model developed in section~\ref{sec:theory}, and especially equation~(\ref{eq:rho}) is valid. As long as this condition is fulfilled, the atomic signal is to first order insensitive to any offset or residual laser frequency noise and cavity length fluctuations, as shown by equation~(\ref{eq:S2}).

With the given experimental parameters, the theoretical shot noise limited resolution for a single scattered photon is
\begin{equation}
	\delta N_\text{th, SNL}  = 3.3 \text{ atoms} \quad \text{for } s = 0 \text{, and } n_\gamma T = 1,
\end{equation}
in principle enabling the entrance into the quantum non-destructivity regime with as little as a few tens of atoms.

\subsection{Experimental results}

The detection system is operated within the experimental cycle by illuminating the cavity with the phase modulated detection beam which has a total power $P_i = $ 4.3~mW and a modulation depth $\beta \simeq 0.31$, such that the heterodyne lock system is robust enough. This gives a total intra-cavity power of $P_c = 4 \eta_\text{cpl}\mathcal{F} J_2^2(\beta)P_i/\pi = $ 10~mW, a saturation parameter $s = P_c/SI_\text{sat} = 8.1\times 10^3$, and a photon scattering rate $n_\gamma = 38$~photons/$\mu$s. The atomic phase shift per atom is $\phi^\text{at}_{+2}/N = -\phi^\text{at}_{-2}/N = 94$~nrad/atom.

\begin{figure}
	\includegraphics[width=\columnwidth]{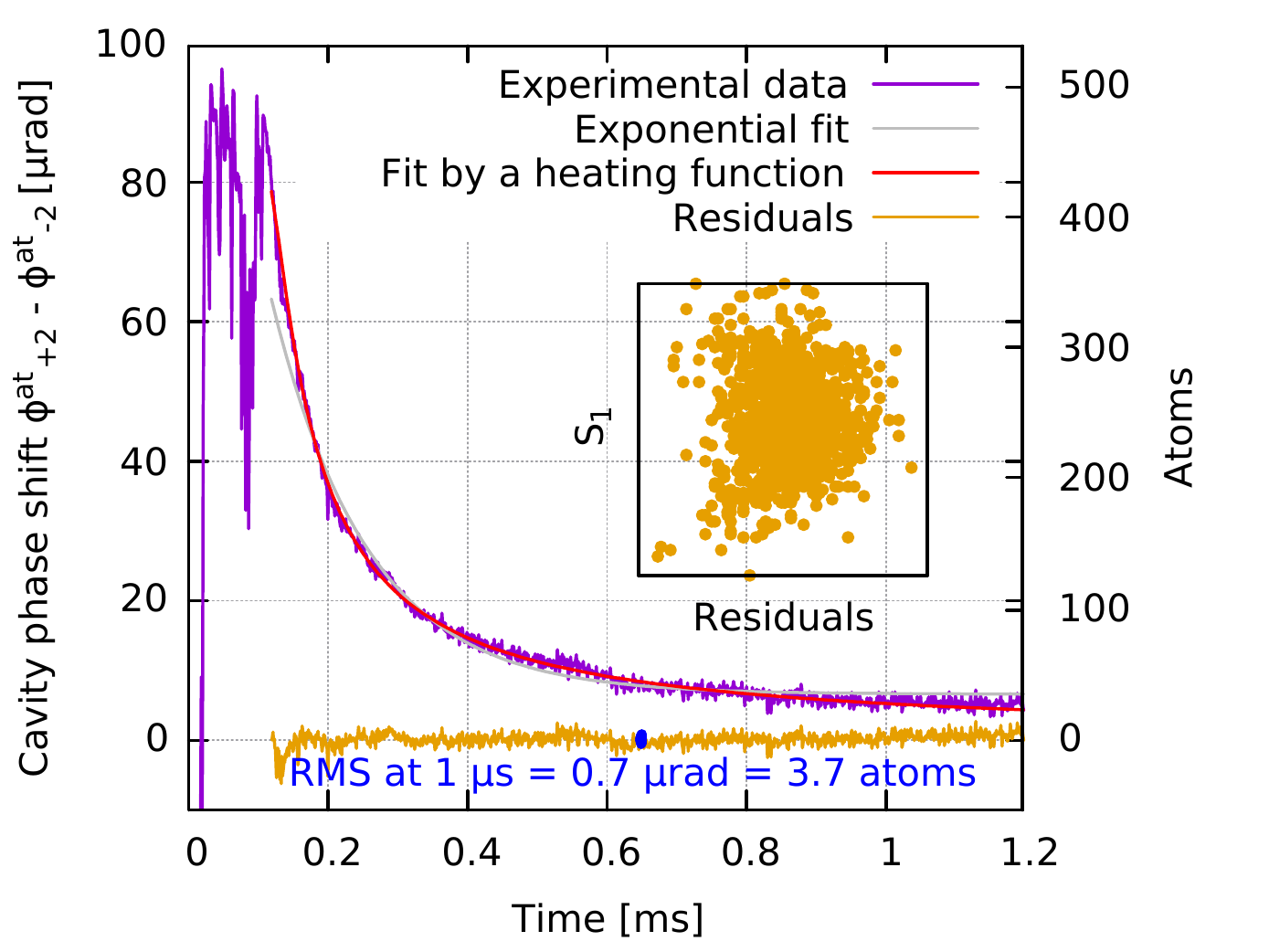}
	\caption{\label{fig:atomdecay}Atomic signal for a trap depth of $U = 750~E_{R,461}$. The sampling rate is 1~MHz. The decay of the trapped atoms is fit by equation~(\ref{eq:heating}) in which a vertical and horizontal offset are added as fit parameters. The scattering rate given by the fit is $n_\gamma = 38$~photons/$\mu$s, in agreement with the value deduced from the optical power injected into the cavity. The grey curve shows that an exponential decay does not fit the experimental data as well as the heating function of equation~(\ref{eq:heating}). The inset shows that the atomic signal $\mathcal{S}_1$ and the residual error signal of the lock of the laser on the cavity are mostly uncorrelated, which confirms that the atomic signal is noise-immune.}
\end{figure}

Figure~\ref{fig:atomdecay} shows an acquisition of the atomic signal with atoms loaded in the optical lattice. At $t = 0$, the detection light is switched on. After a delay of $20~\mu$s, the digital lock is able to stabilise the detection light on the cavity. However, the number of atoms, expected to be a few thousands, introduces a phase shift significantly larger than $\pi/\mathcal{F} \simeq 200~\mu$rad, so that the second order sidebands are no longer simultaneously injected in the cavity. The noise cancellation scheme presented in this paper is therefore not effective, and the atomic signal strongly varies with the laser frequency and cavity length fluctuations. As the atoms scatter photons, they are escape out of the lattice trap. At $t \simeq 150~\mu$s, the number of atoms has sufficiently reduced such that the noise cancellation technique is effective, and the atomic signal becomes noise-immune (see inset of figure~\ref{fig:atomdecay}). The atomic signal then follows a decay that can be fit with a heating function~\cite{Lodewyck2009}:
\begin{equation}
	\label{eq:heating}
	N(t) = N_0\left[1 - \exp\left(-\frac{U/E_{R,461}}{n_\gamma t/3}\right)\right].
\end{equation}
This corresponds to a heating model for which the atoms are confined in the Lamb-Dicke regime in the longitudinal direction, and only acquire momentum in the transverse directions. The standard deviation of the fit residuals gives a resolution of:
\begin{equation}
	\delta N_\text{experimental} = 3.7~\text{atoms},
\end{equation}
for an acquisition time of 1$~\mu$s, during which each atom scatters 38 photons. This shows the classical non-destructivity of the detection scheme, as this number of scattered photons is low enough not to push the atoms out of the lattice. Extrapolating to $n_\gamma T = 1$ yields $\delta N = 23$, enabling the entrance into the quantum regime for $N > 500$ atoms.

The value $\delta N_\text{experimental}$ can be compared to the theoretical shot-noise limited resolution $\delta N_\text{SNL} = 0.7$~atoms, obtained by evaluating equation~(\ref{eq:deltaN}) with the experimental parameters listed above, and $T = 1~\mu$s. An analysis of the dependance of the noise with the optical power reveals that the discrepancy between $\delta N_\text{experimental}$ and $\delta N_\text{SNL}$ is mostly due to electronic noise, and, to a lower extent, to a larger shot-noise than expected. We attribute the latter to extra losses in the photodetector. With an improved photodetector, we can therefore expect a five-fold improvement to the SNR.

\section{Prospects}

The detection scheme demonstrated in this paper offers many perspectives for optical lattice clocks. First, exploiting the high SNR of the detection scheme can help run the clock with a lower number of atoms, and thus a shorter loading time. Figure~\ref{fig:atomdecay} shows that a low detection noise can be reached even with as little as a few hundreds of atoms probed atoms. As an example, we were able to load this amount of atoms with a clock cycle time of 300 ms comprising a 150 ms clock probe pulse. With this sequence featuring a 50\% duty cycle, it would therefore be possible to operate a dead-time free clock using two atomic ensembles~\cite{schioppo2016ultrastable,1367-2630-12-6-065026} even with a clock laser which has a moderate coherence time. Exploiting the classical non-destructivity of the detection will allow us to further improve the duty cycle by recycling the atoms from clock cycle to clock cycle, thus reducing the contribution of the Dick effect to the clock instability.

Several improvements can be envisioned for the scheme presented here. First, the detection system can be extended to detect more atoms by actively adjusting the modulation frequency $\Omega$ so that the two modulation sidebands are coupled into the cavity even when the resonances of the cavity are moved by more than their linewidth. Second, reaching a lower destructivity can be achieved by further reducing the modulation depth, hence coupling less power to the cavity. Because locking the detection laser on the cavity by beating the first and second order sidebands, as proposed here, becomes challenging, another EOM could be used to generate an additional set of first order sidebands with sufficient power to obtain a sizable error signal.

With these proposed improvements, it will be possible to reach the quantum regime, in which entangled states of the atomic ensemble can help reach a sub-QPN frequency stability. The cooperativity of the atom-cavity system presented in this paper, defined by $\eta_0 = 6 \mathcal{F}\lambda^2/\pi^3 w_0^2 = 0.45$ will allow for a significant amount of spin squeezing for the typical atom number of $10^4$ usually operated in OLCs.

\section{Conclusion}

In this paper, we have successfully implemented a cavity-assisted non-destructive detection scheme on an optical lattice clock with strontium atoms. For this, a single phase modulated laser beam is used to lock the laser on the cavity and to detect the atomic signal at the same time, via an heterodyne beatnote. This configuration makes the scheme simple and adapted to operational optical clocks. This detection scheme can measure the number of trapped atoms with a resolution of a few atoms, while scattering a number of photons sufficiently low for the atoms to remain trapped in the optical lattice. This classical non-destructivity will help improve the clock frequency stability by reducing the clock dead-time. Furthermore, simple technical improvements in the detection scheme will make it suitable to reach the quantum non-destructivity regime in which the clock stability can overcome the quantum projection noise limit.

\section*{Acknowledgements}
We acknowledge funding support from Agence Nationale de la Recherche (Labex First-TF ANR-10-LABX-48-01, ANR-16-CE30-0003-01), the European Metrology Research Programme (EXL-01 QESOCAS), the EMPIR 15SIB03 OC18, and Conseil R\'{e}gional \^{I}le-de-France (DIM Nano'K). The EMRP is jointly funded by the EMRP participating countries within EURAMET and the European Union. The EMPIR programme is co-financed by the Participating States and from the European Union's Horizon 2020 research and innovation programme

\end{document}